\input harvmac
\newcount\figno
\figno=0
\def\fig#1#2#3{
\par\begingroup\parindent=0pt\leftskip=1cm\rightskip=1cm
\parindent=0pt
\baselineskip=11pt
\global\advance\figno by 1
\midinsert
\epsfxsize=#3
\centerline{\epsfbox{#2}}
\vskip 12pt
{\bf Fig. \the\figno:} #1\par
\endinsert\endgroup\par
}
\def\figlabel#1{\xdef#1{\the\figno}}
\def\encadremath#1{\vbox{\hrule\hbox{\vrule\kern8pt\vbox{\kern8pt
\hbox{$\displaystyle #1$}\kern8pt}
\kern8pt\vrule}\hrule}}

\overfullrule=0pt

\Title{MIT-CTP-2546, TIFR-TH/96-19}
{\vbox{\centerline{Comparing decay rates for black holes and  D-branes}}}
\smallskip
\centerline{Sumit R. Das\foot{E-mail: das@theory.tifr.res.in}}
\smallskip
\centerline{\it Tata Institute of Fundamental Research}
\centerline{\it Homi Bhabha Road, Bombay 400 005, INDIA}
\smallskip
\centerline{and}
\smallskip
\centerline{Samir D. Mathur\foot{E-mail: me@ctpdown.mit.edu}}
\smallskip
\centerline{\it Center for Theoretical Physics}
\centerline{\it Massachussetts Institute of Technology}
\centerline{\it Cambridge, MA 02139, USA}
\bigskip

\medskip

\noindent

We compute the leading order (in coupling) rate of emission of low
energy quanta from a slightly nonextremal system of 1 and 5 D-branes.
We also compute the classical cross-section, and hence the Hawking
emission rate, for low energy scalar quanta for the black hole
geometry that corresponds to these branes (at sufficiently strong
coupling).  These rates are found to agree with each other.

\Date{June, 1996}

\def\cA{{\cal A}}

\newsec{Introduction}

Recently D-branes
\ref\dbrane{For a review and references to the original literature, 
see J. Polchinski, S. Chaudhuri and C. Johnson,
hep-th/9602052.} have provided significant insights into the thermodynamics
of black holes in string theory \ref\horowitz{For a review of black hole
entropy in string theory and references, see G. Horowitz, gr-qc/9604051.}.
In particular, a configuration of 1-D branes
and 5-D branes compactified on a suitable five dimensional compact space
(which can be $K3 \times S^1$ or $T^5$) represents a six dimensional
black string, or a five dimensional black hole for distance scales larger
than the size of the string. For such a configuration which is BPS
saturated, it was shown that the degeneracy of states for a given mass
and charges leads to an entropy which is {\it exactly} equal to the
Beckenstein-Hawking entropy of the corresponding black hole 
\ref\stromvafa{A. Strominger and C. Vafa,
hep-th/9601029}. This is a realization of the idea that 
the degeneracy of string states
is responsible for Hawking-Beckenstein entropy 
\ref\susskind{L. Susskind, hep-th/9309145;
J. Russo and L. Susskind, {\it Nucl. Phys.}~{\bf B437} (1995) 611},
which was pursued for holes with NS-NS charges and 
zero extremal area in \ref\sen{A. Sen,
{\it Nucl. Phys.}~{\bf B440} (1995) 421 and {\it Mod. Phys. Lett.}
~{\bf A10} (1995) 2081.} and through 
quantisation of metric fluctuations for nonzero area holes in  
\ref\wilczek{F. Larsen and F. Wilczek, 
hep-th/9511064}. For other work on extremal 
holes, see \ref\nsns{ A.A. Tseytlin,
{\it Mod. Phys. Lett.}~{\bf A11}(1996) 689, hep-th/9601177; 
M. Cvetic
and A.A. Tseytlin, {\it Phys. Rev.}~{\bf D53} (1996) 5619;
J.C. 
Beckenridge, R.C. Myers, A.W. Peet and C. Vafa, hep-th/9602065; J. Maldacena
and A. Strominger, hep-th/9303060; C.V. Johnson, R.R. Khuri and
R.C. Myers, hep-th/9603061; A.A. Tseytlin, hep-th/9603099; R. Dijkgraaf,
E. Verlinde and H. Verlinde, hep-th/9603126; J. Breckenridge,
D. Lowe, R. Myers, A. Peet, A. Strominger and C. Vafa, hep-th/9603078;
I. Klebanov and A. Tseytlin, hep-th/9604166; F. Larsen
and F. Wilczek, hep-th/9604134; 
V. Balasubramaniam and F. Larsen, hep-th/9604189;
A.A. Tseytlin, hep-th/9605091.}. 

The nature of non-BPS states of D-branes was discussed for the
D-string in \ref\dasmathur{S.R. Das and S.D. Mathur,
hepth/9601152.}. Remarkably, it was found in
\ref\callanmalda{C.G. Callan and J. Maldacena, hep-th/9602043} and
\ref\horostrom{G. Horowitz and A. Strominger, hep-th/9602051} 
that the degeneracy of
slightly excited non-BPS states also agrees with the
Beckenstein-Hawking entropy for the corresponding nonextremal
hole. The nature of excitations responsible for the nonextremal
entropy is in fact quite similar to that of the D-string - these are
now open strings moving along the 1 brane, but with polarizations
which lie entirely in the brane directions \foot{The nature of these
open strings depend on whether the $5$ brane charge is equal to or
greater than 1. See \ref\hormalstr{G. Horowitz, J. Maldacena and
A. Strominger, hep-th/9603109.}}. In \ref\klebanov{S.S. Gubser, I. Klebanov
and A.W. Peet, hep-th/9602135; I. Klebanov and A. Tseytlin,
hep-th/9604089 } it was found that
the non-BPS excitations of some other black branes have
thermodynamical properties that agree with predictions from black hole
thermodynamics, though there are also many exceptions that arise at
least from a naive application of this correspondence. See \ref\nonext{M.
Cvetic and D. Youm, hepth/9603147, hep-th/9605051, hep-th/9606033;
G. Horowitz, D. Lowe and J. Maldacena, hep-th/9603195; M.Cvetic and
A. Tseytlin, hep-th/9606033} for some recent work on nonextremal holes.

Even more remarkably, \callanmalda\ showed that the rate of decay of
such an excited state via annihilation of oppositely moving open
strings into a closed string which then escapes from the brane
qualitatively (i.e. apart from numerical factors)
agrees with the expectations from hawking radiation
and the effective temperature of the outgoing state agrees exactly
with the hawking temperature of the nonextremal hole. 
(The analogue of this calculation for absorption has been
 recently analysed in \ref\spenta{A. Dhar, G. Mandal and S.R. Wadia,
hep-th/9605234.} where it is also shown that the classical 
cross-section for absorption of low energy scalars by the black hole
is proportional to the area like the D-brane calculation in
\callanmalda. )

These results for nonextremal holes could appear surprising, since the
regime in which the branes act as black holes is when the open string
coupling is large, whearas the above calculations were performed at
weak coupling. (It was argued in \dasmathur, however, that when the
length of a D-string is large, the weak coupling results remain
accurate for long wavelength modes even at somewhat larger coupling.)
The issues in this regard are not clear, and there is no clear reason
to expect that non-BPS processes agree between D-brane calculations at
low orders in coupling and semiclassical results based on the geometry
produced by the branes at stronger coupling.

In this paper we carry out two calculations:

\item{(a)}  
In the first calculation we take the configuration discussed in
\stromvafa\ and \callanmalda\ .  We take the case with a single
5-brane wound on a $T^5$, and a collection of D-strings (wound along
one of the directions of the $T^5$), with net momentum along the
D-string direction. We compute the rate of emission of low energy
quanta from a slightly nonextremal state of this configuration. To
compute the decay amplitude we use the Born-Infeld action for
describing the long wavelength fluctuations of the D-string within the
5-D-brane.  (Note that as discussed in \dasmathur\ and
\ref\susskindmalda{J. Maldacena and L. Susskind,
hep-th/9604042} the D-strings should behave as one long string rather
than a collection of singly wound strings, and we adopt this model
here.)  From this action we derive the coupling of the oscillations of
the string to the gravitons in the 10-dimensional spacetime. Because
of the compactification on $T^5$, we further decompose these gravitons
into scalars, vectors and gravitons of the 5-dimensional
non-compact spacetime . Because the D-string can only vibrate within
the 5-brane,  of these fields only  the 5-dimensional scalars
will be emitted.

\item{(b)} 
In the second calculation we compute the probability that a low energy
scalar is absorbed by an extremal black hole with geometry
corresponding to the charges carried by the D-branes in calculation
(a) above. Here we follow the method of \ref\unruh{
W.G. Unruh, {\it Phys. Rev.}~{\bf D14} (1976) 3251.}, which has also
been used in the context of a slightly nonextremal 5-dimensional hole
recently in \spenta . It is found that the effective
cross-section is the area of the horizon, just as was the case in the
3+1 dimensional case of \unruh .  At least in 3+1 dimensions the
absorption cross section (and therefore the emission rate) of low
energy particles of spin one and spin two vanish for low energies
\ref\page{D. Page, {\it Phys. Rev.}~{\bf D13}(1976) 198.}. It
is plausible that such is the case here as well, but we do not address
that calculation in this paper.

We find that the absorption rate for scalars agree between the
calculations (a) and (b). Photons and gravitons are not emitted at
this order in the calculation (a); if the 3+1 dimensional calculation
is an accurate guide to the 4+1 dimensional extremal case then there
is agreement here as well.

\newsec{General features of D-string amplitudes}

In a companion paper \ref\dasmathurtwo{S.R.Das and S. Mathur,
to appear} we have considered
various aspects of amplitudes for the decay of an excited D-string
into a massless closed string state of low energy. Since the relevant
massless excitations of the 1D brane and 5D brane configuration are
similar to those of a D-string, we recall some of the basic features.
For such low energy processes we can use the Dirac-Born-Infeld action
to compute the amplitudes. This may be written in terms of
the coordinates of the D-string $X^\mu(\xi^m)$
(where $\mu$ runs over all the $10$ indices whereas $\xi^m$ are
parameters on the D-string worldsheet) and the gauge fields
on the D-string worldsheet $A^m (\xi^m)$ as follows
\ref\broninfeld{See e.g. Ref [1]; A. Tseytlin, hep-th/9602064.}
\eqn\throne{S_{BI} = T\int d^2\xi~e^{-\phi(X)}
{\sqrt{{\rm det}~ [G_{mn}(X) + B_{mn}(X) + F_{mn}]}}}
where $F_{mn}$ denotes the gauge field strength on the D-string worldsheet
and $G_{mn}, B_{mn}$ are the background (string) metric and NS-NS 2-form fields
induced on the worldsheet
\eqn\thrtwo{G_{mn} = G^{(S)}_{\mu\nu}(X)\partial_m X^\mu \partial_n X^\nu
~~~~~~~~~~~B_{mn} = B_{\mu\nu}(X)\partial_m X^\mu \partial_n X^\nu}
$T$ is a tension related to the D-string tension by
$T^D=e^{-\phi/2}T$.
We will work in the static gauge which means
\eqn\thrthree{X^0 = \xi^0~~~~~~~~~~~X^1 = \xi^1}
In this gauge the massless open string
fields which denote the low energy excitations of the brane are
the transverse coordinates $X^i(X^0,X^1),~~i = 2,\cdots 9$. 

In the following we will set the gauge field and the RR field to be zero.
The lowest order interaction between the metric fluctuations around
flat space and the open string modes is obtained by expanding the
metric as $G_{\mu\nu} = \eta_{\mu\nu} + 2\kappa h_{\mu\nu} (X)$,
expanding the transverse coordinates $X^i (\xi)$ around the brane
position $X^i = 0$ and treating $h_{\mu\nu}$ and $X^i$ to be
small. Here $\kappa$ is the ten dimensional gravitational coupling
present in the bulk action (in terms of the einstein metric)
\eqn\decone{S = {1 \over 2\kappa^2}\int d^{10}x~{\sqrt{g}}[R - 
{1\over 2} (\nabla \phi)^2 + \cdots]}
For purely transverse gravitons, i.e. only $h_{ij} \neq 0$, the terms
upto two open string fields is (after rescaling $X \rightarrow
{\sqrt{T^D}}X$)
\eqn\thrnine{{1\over 2}(\delta_{ij} + 2\kappa h_{ij})\partial_\alpha X^i
\partial^\alpha X^j}
Note that to this order the lagrangian is independent of $T^D$.
We will be interested in the decay amplitude for a closed string state
in a specific polarization state, say $h_{67}$. Note that the quadratic
action for the $h_{ij}$ which follows from the bulk action \decone\ is
(in the harmonic gauge)
\eqn\dectwo{\int d^{10} x {1\over 2} (\partial h_{ij})(\partial h_{ij})}
Since $h_{ij}$ are symmetric the field $h_{67}$ does not have a properly
normalized kinetic term. The properly normalized field is then
${\bar h}_{67} = {\sqrt{2}}h_{67}$. This means that the interaction term
with a ${\bar h}_{67}$ with two open string fields is, from \thrnine,
\eqn\decthree{{\sqrt{2}}\kappa~ {\bar h}_{67} \partial X^6 \partial X^7}

Consider a D-string which is excited above the BPS state by addition
of a pair of open string states with momenta (on the worldsheet)
$(p_0,p_1)$ and $(q_0,q_1)$ respectively. The decay of this
state into the extremal state is given by the process of annihilaton
of this pair into a closed string state, like a graviton. For a
graviton represented by ${\bar h}_{ij}$
with momentum $(k_0, k_1, {\vec k})$ (where ${\vec k}$ denotes the
momentum in the transverse direction), the
leading term for this amplitude for low graviton energies can be
read off from \decthree\ as
\eqn\thrten{\cA_D = {\sqrt 2}\kappa A_D = {\sqrt{2}}\kappa  p \cdot q}
When the outgoing graviton does not have any momentum along the
string direction one has
\eqn\threleven{p \cdot q = p^0q^0 - p^1q^1 = 2 |p_1|^2}
where we have used momentum conservation in the string direction
and the masslessness of the modes. As shown in \ref\hashikleba{A. Hashimoto
and I. Klebanov, hep-th/9604065.} a direct conformal field theory
calculation of the decay rate agrees with the above answer for low
energies and transverse polarizations.

The pair of  colliding open strings is  part of a one dimensional gas of 
open strings. To
obtain the decay rate of the nonextremal state one has to compute
the decay rate for this specific initial state and then average
over initial states. The latter averaging is responsible for
the thermal nature of the outgoing closed string.

\newsec{D-brane thermodynamics}

In this section we discuss the thermodynamics for massless
open string states moving on the D-brane.
We essentially have the statistical mechanics of massless
particles on the brane. Typically
we have a net momentum in one of the directions so actually we
have an ensemble with given total energy $E$ and total
momentum $P$ along say the $X^1$ direction. For large size
of the D-brane we can approximate this by a canonical type
ensemble characterized by an inverse momentum $\beta$ and
a chemical potential $\alpha$ as follows. Let there be
$n_r$ particles with energy $e_r$ and momentum in the $X^1$
direction $p_r$. Define a partition function ${\cal Z}$ by
\eqn\foone{{\cal Z} = e^h = \sum_{states} {\rm exp}~[ -\beta\sum_r
n_r e_r - \alpha\sum_r n_r q_r]}
Then $\alpha, \beta$ are determined by requiring
\eqn\fotwo{ E = -{\partial h \over \partial \beta}~~~~~~~~~
P = -{\partial h \over \partial \alpha}}
The average number of particles $n_r$ in state $(e_r, p_r)$ is
then given by
\eqn\fothree{ \rho (e_r, p_r) = {1 \over e^{\beta e_r + \alpha p_r}
\pm 1}}
where as usual the plus sign is for fermions and the minus sign is
for bosons. Finally the entropy $S$ is given by the standard thermodynamic
relation
\eqn\fofour{ S = h + \alpha P + \beta E}

For the case of a D-string with $f$ species of bosons and $f$
species of fermions the above quantities may be easily evaluated
\eqn\fofive{\eqalign{&P = {fL\pi\over 8}[{1\over (\beta + \alpha)^2}
-{1 \over (\beta - \alpha)^2}]~~~~~~~~~~~
E = {fL\pi\over 8}[{1\over (\beta + \alpha)^2}
+{1 \over (\beta - \alpha)^2}]\cr
&~~~~~~~~~~~~~~~~~~~~~~S = {fL\pi \over 4}[{1\over \beta + \alpha}
+{1 \over \beta - \alpha}]}}
Since we have massless particles in one spatial dimension, they
can be either right moving, with $e_r = p_r$ or left moving
$e_r = -p_r$. The distribution functions then become
\eqn\fosix{\eqalign{&\rho_R = {1 \over e^{(\beta + \alpha)e_r}\pm 1}~~~~
~~~~~~~
{\rm R}\cr
&\rho_L = {1 \over e^{(\beta - \alpha)e_r} \pm 1}~~~~~~~~~~~
{\rm L}}}
Thus the combinations $T_R = 1/(\beta + \alpha)$ and 
$T_L = 1/(\beta - \alpha)$
act as {\it effective} temperatures for the right and left moving
modes respectively. In fact all the thermodynamic quantities
can be split into a left and  a right moving piece : 
$E = E_R + E_L~~~~P = P_R + P_L~~~~~S = S_R + S_L$ in an obvious
notation. The various quantities $E_L, E_R, P_L, P_R, S_L, S_R$
may be read off from \fofive. 
>From \fofive\ we get
\eqn\foseven{ T_R = {\sqrt{8 E_R \over L\pi f}}
~~~~~~~~~~~~~T_L = {\sqrt{8 E_L \over L\pi f}}}
The extremal state corresponds to $P_L = E_L = 0$ so that $E = P$.
Finally we note that 
\eqn\foeight{ T_L = {4 S_L \over \pi f L}~~~~~~~~~~~~~~
T_R = {4 S_R \over \pi f L}}
This relation will be crucial in physical properties of the
decay rate which we now calculate.

Finally we note that the left right splitting is a feature of
statistical mechanics of massless particles in one (spatial) dimension,
which is relevant to the D-string as well as other examples of
D-strings bound to 5D branes to be considered below. For higher
dimensional branes, like the three brane, such simplifications
may not apply, except for certain limiting values of the parameters. 
However the method of introducing a chemical
potential for the total momentum can be used in this case as well,
as we will discuss in a later communication.

\newsec{The decay rate for D-string}

The S-matrix element for the decay is given by
\eqn\fione{S_{fi} = {\sqrt 2}\kappa (2\pi)^2 \delta (p_0+q_0-k_0)
\delta(p_1+q_1-k_1) {(-i A_D) \over {\sqrt{(2p_0 L)(2q_0 L)
(2 k_0 V_9)}}}}
where $V_9$ denotes the nine dimensional spatial volume and
$L$ is the length of the D-string. If $R$ is the radius of
compactification of the string direction, then for a multiply
wound string one has $L = 2\pi n_w R$. The total spatial
volume $V_9 = V_8 (2\pi R)$ where $V_8$ is the volume of the
noncompact space. The decay rate for this pair to produce 
a graviton of zero momentum in the string direction and with
a {\it given} polarization is then
\eqn\fitwo{\Gamma(p,q,k) = {\kappa_9^2 (2\pi)^2 \over 4L}
\delta (p_0+q_0-k_0)\delta(p_1+q_1-k_1)
{|A_D|^2 \over p_0 q_0 k_0 V_8 }{V_8 [d^8 k]
\over (2\pi)^8}}
where we have introduced the nine dimensional gravitational
coupling $\kappa_9^2 = {\kappa^2 \over 2\pi R}$.

To obtain the total rate to produce a graviton with the given
momentum we have to average over all initial states. This includes
a sum over the momenta and polarizations of the open string
states.
When the open string momentum quantum
numbers are large (but still much smaller than the net
momentum quantum number) this means that we have to
multiply by the relevant thermal distribution functions
obtained in the previous section. Denoting the distribution
functions by $\rho (p_0,p_1)$ and $\rho(q_0,q_1)$ the total
decay rate is
\eqn\fithree{\Gamma (k) = \int_{-\infty}^{\infty} {Ldp_1 \over 2\pi}
\int_{-\infty}^{\infty} {L dq_1 \over 2\pi}~\Gamma (p,q,k) 
\rho (q_0, q_1) \rho (p_0,p_1)}

We will consider the emission of a graviton of a specified
polarization, say $\epsilon_{67} = 1$ and the others zero.
Let us first specialize to the case where the outgoing particle
has $k_1 = 0$. Then this graviton may be produced only by
the annihilation of two open strings with exactly equal and
opposite momenta. There are two possible polarizations of
the initial state. The first would be when the left moving open
string has a polarization in the 6 direction and the right moving
in the 7 direction and the second would be the other way round.
However once the momentum integral in \fithree\ is over the entire
range $[-\infty,\infty]$ we have automatically summed over these
two polarization states. 

We will evaluate the rate for
low energies where we can use the expression for ${\cal A}_D$ in
the previous section.
The integral over $q_1$ sets $q_1 = - p_1$ and one is left with
\eqn\ffione{\Gamma (k) = {\kappa_9^2 L [d^8 k] \over  k_0 (2\pi)^8}
\int_{-\infty}^\infty dp_1 \delta (2|p_1| - k_0) |p_1|^2 \rho(|p_1|,p_1)
\rho (|p_1|,-p_1)} 
The integration can be done
\eqn\sifive{\eqalign{&\int_{-\infty}^\infty \delta (2 |p_1| - k_0) 
|p_1|^2~\rho(|p_1|,p_1)~\rho (|p_1|,-p_1)\cr
&= 2 \cdot {1\over 2} ({k_0 \over 2})^2 \rho({k_0 \over 2},{k_0 \over 2})
\rho({k_0 \over 2},-{k_0 \over 2})}}
where the factor 2 comes from the two values of $p_1$ where the
delta function clicks and the factor of ${1\over 2}$ comes from
the Jacobian involved in performing the integration.
Note that the two distributions which appear are the left and right
distributions defined in \fosix.

Since the intial state is only slightly nonextremal, we have
$E_L = E_{BPS} + \Delta E$ and $E_R = \Delta E$ and 
$\Delta E << E_{BPS}$
Then \foseven\ implies that ${p_0 \over T_L}$ is small.

First consider the case where the open string states are
bosonic. Then we can approximate $\rho_L (p_0)$ in  \ffione\ as
\eqn\fifour{\rho_L (p_0) \sim {T_L \over p_0}= {4 S_L \over p_0 L\pi f}}
where we have used \foeight. In this case of a multiply wound 
D-string we have $f = 8$. 
The final result is
\eqn\fifive{\Gamma (k) = {\kappa_9^2 S_{ext} 
\over 2^{10} \pi^9 } [d^8 k] { 1 \over e^{k_0 \over T_{eff}}-1}}
where we have defined an effective temperature 
of the graviton spectrum to be
\eqn\fisix{ T_{eff} = 2 T_R}
and have used the fact that $S_L \sim S_{ext}$ where $S_{ext}$ is 
the extremal entropy.
It is important to note that the size of the compact direction
has disappeared from this answer.

When the initial open string states are fermionic $\rho_L (p_0)$
becomes order unity in the near extremal case rather than the
large quantity \fifour\ for the bosonic case. Thus when the
extremal entropy is large, these states do not contribute to
 $\Gamma (k)$ in the leading order.

The crucial fact about \fifive\ is that the answer comes
out to be proportional to the extremal entropy. This is a consequence
of the fact that we have essentially one dimensional thermodynamics
where the temperature is proportional to the entropy. In the
next section we will deal with brane configurations which correspond
to nonzero extremal horizon area and where the extremal entropy
agrees with the Hawking-Beckenstein formula and hence proportional
to the horizon area. We then get a result proportional to the horizon
area.

\newsec{Decay Rate for 1-Brane 5-brane configurations}

The results of the previous section may be used to calculate the
decay process for situations where the configuration of branes
produce a spacetime which has a large horizon in the extremal
limit. In the following we will do so for the model similar to
that considered in \callanmalda.
We will consider a configuration of one 5D-brane wrapped around
a $T^5$ in the $(X^5-X^6-X^7-X^8-X^9)$ direction and single
D-string wound $Q_1$ times around the $X^5$ direction. The
radius of the $X^5$ direction is $R$ which is taken to be large.
When the $T^4$ in the $(6-7-8-9)$ direction is small this
represents a black string in six dimensions one of which is
compact but large. 

In this case, as argued in \hormalstr\
the low energy excitations of the
system are described by massless modes of open strings which begin and
end on the D-string and whose polarization vectors $\lambda^i$ lie in
the $6-7-8-9$ plane. These modes thus live on an effective length $L=
2\pi Q_1 R$.  There are $4$ such bosonic and $4$ fermionic modes. The
extremal state corresponds to the case when all these open strings are
moving in the same direction and a nonextremal situation corresponds
to strings moving in either directions.

We thus have a situation similar to the case of a single D-string
discussed in the previous sections with the number of flavors
$f$ in Section 5 being $4$. As pointed out in
\susskindmalda\
the thermodynamic formulae are valid when
extensivity holds. This is indeed the case here in the
``fat hole'' limit. For example
in the extremal limit $E_R = 0$ and $E_L = {N \over R}$ and
from \foseven\ one has
\eqn\sione{T_L L = 2 {\sqrt{N Q_1}}}
which is large when $Q_1$ and $N$ are of the same order
and large. The extremal entropy is
\eqn\sitwo{S_{ext} = 2\pi{\sqrt{NQ_1}} = {A_H \over 4 G_5}}
where $A_H$ is the horizon area and $G_5$ is the five dimensional
Newton constant.

The amplitude for the decay of an nonextremal hole is identical to
that in section 4. However since the low energy open string modes
have polarizations $\lambda^i$ where $i = 6 \cdots 9$ 
it follows from \thrten\ that
at lowest
order the only gravitons which are produced have polarization
in these directions. From the five dimensional point of view
these are in fact scalars. In the following we will examine the
decay rate for the production of a given polarization state,
say the situation in which only $\epsilon_{67}$ is nonzero.

The S-matrix element is identical to \fione. However the decay
rate $\Gamma (p,q,k)$ for an outgoing closed string mode
which has no momentum along the $X^5$ direction is a slight
modification of \fitwo\
\eqn\sithree{\Gamma(p,q,k) = {\kappa_5^2 (2\pi)^2 \over 4L}
\delta (p_0+q_0-k_0)\delta(p_1+q_1-k_1)
{|A_D|^2 \over p_0 q_0 k_0 V_4 }{V_4 [d^4 k]
\over (2\pi)^4}}
$V_4$ denotes the volume of the spatial noncompact four
dimensions, while $\kappa_5^2 = {\kappa^2 \over 2\pi R {\tilde V_4}}$
where ${\tilde V_4}$ denotes the volume of the compact directions
$X^6 \cdots X^9$. The total decay rate is given by
\fithree ,
the $\Gamma(p,q,k)$ in this equation being given by \sithree.

We now use the low energy result for ${ A}_D$ given in \thrten\
and \threleven\ and integrate over $q_1$ and $p_1$. These integrations
are identical to those in the previous section and one has
\eqn\siione{\Gamma (k) = {\kappa_5^2 L [d^4k] \over (2\pi)^4
k_0}~({k_0 \over 2})^2~\rho_L(k_0/2)\rho_R (k_0/2)}
For low energies we use \fifour\ with $f = 4$ to get
\eqn\siten{\rho_L(p_0) \sim {T_L \over p_0} = {A_H \over 4 \pi G_5 L p_0}}
Plugging \siten\ into \siione, and performing the $p_1$ integral
as in the previous section and using $\kappa_5^2 = 8\pi G_5$
we finally get
\eqn\sisix{\Gamma (k) = {A_H \over 16\pi^4} [d^4k] \rho_R(k_0/2)}
Then the total energy emitted in an energy range $k_0, k_0 + dk_0$
per unit time is obtained by multiplying \sisix\ with $k_0$
and writing out the phase space factor $[d^4k] = 2\pi^2 k_0^3 dk_0$.
We finally obtain
\eqn\sisix{ {dE(k) \over dt} = {A_H \over 8 \pi^2}
{k_0^4 dk_0 \over e^{\beta_H k_0} - 1}}
The Hawking temperature is $T_H = 1/\beta_H = 2 T_R$ and as noted
in \callanmalda\ agrees with the temperature defined by the surface
gravity at the horizon.

\newsec{The classical cross section.}

In this section we find the classical cross section for absorption of
low energy scalar quanta into the extremal black hole having the
charges of the D-brane model discussed in the previous section.  The
method of computing such low energy cross-sections 
for four dimensional holes was given in \unruh,
and a calculation for a slightly nonextremal hole in five dimensions
has been recently carried out in \spenta.
We will carry out the
calculation for the extremal hole, and observe that it agrees with the
extremal limit of \spenta, though the actual details of the
calculation differ in the extremal case and in the slightly
nonextremal case. 

The extremal metric for the five dimensional black hole is given by
\eqn\tenone{ds^2 = - [f(r)]^{-{2\over 3}} dt^2 +[f(r)]^{{1\over 3}} dr^2
+[f(r)]^{{1\over 3}} r^2 d\Omega_3^2}
where
\eqn\tentwo{ f(r) = (1 + {Q_1 \over r^2})(1 + {Q_2 \over r^2})
(1 + {Q_3 \over r^2})}

The massless minimally coupled scalar wavefunction is
\eqn\tenthree{\phi(r,t)=R(r)e^{-i\omega t}}
where we restrict to spherically symmetric wavefunctions, since as
shown in \unruh, the higher angular momentum components are not
absorbed in the limit of low frequencies. The wave equation reduces to
\eqn\tenfour{[{d^2\over dr^2}+\omega^2 f(r)-{3\over 4r^2}]\psi(r)=0}
 where
\eqn\tenfive{\psi(r)=r^{3/2} R(r)}

The idea of \unruh\ is to solve this equation approximately in three
regions, and match the solutions across the boundaries of the regions.
It is assumed that it is adequate to keep only the lowest order terms
in $\omega$ in each region; we will adopt this assumption here as
well.

\subsec{ Outer region:} 

The outermost region is $r>>Q_i^{1/2}$. In this region we get 
\eqn\tensix{f=1+{Q\over r^2}}
where
\eqn\tenseven{Q=Q_1+Q_2+Q_3}
Defining $\rho=\omega r$, the equation is
\eqn\teneight{[{d^2\over d\rho^2}+(1+{Q\omega^2-3/4\over \rho^2})]\psi=0}
The solution may be written as
\eqn\tenine{ \psi (\rho) = \alpha F(\rho) + \beta G(\rho)} where the
two independent solutions may be written in terms of Bessel functions
\eqn\tenten{\eqalign{&F=\sqrt{{\pi\over 2}}\rho^{1/2}J_{(1-Q\omega^2)^{1/2}}
(\rho)\cr
& G=\sqrt{{\pi\over 2}}\rho^{1/2}J_{-(1-Q\omega^2)^{1/2}}(\rho)}}
In the region $\rho>>1$ we have
\eqn\teneleven{\eqalign{&F=\cos(\rho-\pi/2(1-Q\omega^2)^{1/2}-\pi/4)\cr
& G=\cos(\rho+\pi/2(1-Q\omega^2)^{1/2}-\pi/4)}}
In terms of a shifted coordinate $\rho' = \rho - {\pi \over 4}$ we get
\eqn\tentwelve{\psi=e^{i\rho'}[{i\over 2}(-\alpha e^{i\pi/4 Q\omega^2} 
+\beta e^{-i\pi/4 Q\omega^2})]+e^{-i\rho'}[{i\over 2}
(\alpha e^{-i\pi/4 Q\omega^2} -\beta e^{i\pi/4 Q\omega^2})]}
implying a reflection coefficient
\eqn\tenthirteen{{\cal R}=-e^{i{\pi\over 2} Q\omega^2}
{1-{\beta\over \alpha}e^{-i{\pi\over 2} Q\omega^2}\over
1-{\beta\over \alpha}e^{i{\pi\over 2} Q\omega^2}}}
This  will give for the absorption probablility
\eqn\tenfourteen{ |{\cal A}|^2 = 1 - |{\cal R}|^2}
We thus need to find $\alpha$, $\beta$, from the requirement that only
an inward moving wave exists at the horizon; we will compute these
parameters below.

>From the behavior of the Bessel functions at small argument we find 
that for $\rho<<1$, the solution to the wave equation is
\eqn\tenfifteen{R(r)\sim 
{\sqrt{{\pi \over 2}}}\omega^{{3\over 2}}[{1\over 2}\alpha 
+ {\beta Q \over r^2}]}

\subsec{Intermediate region}

This region has $r\sim Q_i^{1/2}$. Then the equation (for $Q\omega^2<<1$)
is
\eqn\tensixteen{{1\over r^3}{d\over dr}r^3 {dR\over dr}=0}
which gives
\eqn\tenseventeen{R=C+{D\over r^2}}

\subsec{Near horizon region:}

This region is $r<<Q_i^{1/2}$. Here the wave equation is
\eqn\teneighteen{{1\over r^3}{d\over dr}r^3 {dR\over dr}+
{\omega^2 P\over r^6}(1+\mu r^2)R=0}
where $P=Q_1Q_2Q_3$ and $\mu=1/Q_1+1/Q_2+1/Q_3$.  Define
\eqn\tennineteen{u=-{1\over 2r^2}, ~~\rho=u\omega\sqrt{P}, 
~~\eta={1\over 4}\mu\omega\sqrt{P}
={1\over 4}\omega\sqrt{Q_1Q_2Q_3}(1/Q_1+1/Q_2+1/Q_3)}
Then the  wave equation is
\eqn\tentwenty{{d^2R\over d\rho^2}+(1-{2\eta\over \rho})R=0}
The two independent solutions are given by Coulomb functions
\ref\abramo{M. Abramowitz and I. Stegun, {\it Handbook of Mathematical
Functions}}. 
Very close to the horizon, $\rho \rightarrow -\infty$ the two independent
solutions are
\eqn\tentwoone{\eqalign{&F_0=\sin[\pi/4+\rho+\eta\log({\eta\over 2\rho})]\cr
&G_0=\cos[\pi/4+\rho+\eta\log({\eta\over 2\rho})]}}
The particular linear combination we have to pick is
\eqn\tentwotwo{R(r) = G_0(r) -iF_0(r)}
to get an ingoing wave at large negative $\rho$.  At small $|\rho|$ we
 get 
\eqn\tentwothree{R \sim {1\over \sqrt{ 2}}[(1-i)-(1+i)\rho]=
{1\over \sqrt{ 2}}[(1-i)+(1+i){\omega\sqrt{P}\over 2r^2}]}

\subsec{Matching the solutions}

Comparing \tentwothree\ with \tenseventeen\ one gets
\eqn\tentwofour{C={1\over \sqrt{ 2}}(1-i)~~~~
D={1\over 2\sqrt{ 2}}(1+i)\omega\sqrt{P}}
Matching the solution \tenfifteen\ with \tenseventeen\ and using
\tentwofour\ we get
\eqn\tentwofive{\alpha={2(1-i)\over \omega^{3/2}\sqrt{\pi}}~~~~~
\beta={(1+i){\sqrt{P}}\over 2{\sqrt{\pi}}Q \omega^{{1\over 2}}}}
We can now easily compute the absorption probability $|A|^2$ from
\tenfourteen,
\eqn\tentwosix{|A|^2 = {1\over 2} \pi \omega^3 {\sqrt{P}}}
This  may be
expressed in terms of the area of horizon 
$A_H = 2\pi^2 \sqrt{P}$ as
\eqn\tentwoseven{ |A|^2 = {1 \over 4\pi}\omega^3 A_H}

\subsec{The absorption cross section}

We now calculate the absorption cross section of a plane wave
incident on the hole. Let us expand a plane wave as follows
\eqn\tentwoeight{ e^{i\omega z} = K {e^{-i\omega r}\over r^{{3\over 2}}}Z_{000}
+ {\rm {other ~ terms}}}
where we have kept only the radially inward momentum component 
and ignored modes that are not spherically symmetric.  
(As mentioned above, we expect that the higher angular momentum 
waves have an absorption coefficient that is suppressed at low energies.)
  The quantity $Z_{000}$ is
normalized over a three sphere $Z_{000} = {1 \over {\sqrt{2\pi^2}}}$

Once $K$ is known the absorption cross-section of the S-wave
is given by
\eqn\tentwonine{\sigma = |K|^2 |A|^2}

An efficient way of determining $K$ is to integrate both sides of
\tentwoeight\ over all angles with the standard measure of a three
sphere. This projects out only the spherically symmetric part,
thus the first term of the right hand side of \tentwoeight. One gets
\eqn\tenthirty{4 {2 \pi^{{3\over 2}}\over \omega r} \Gamma ({3\over 2})
J_1 (\omega r) = {\sqrt{2\pi^2}} K {e^{-i\omega r} \over r^{{3\over 2}}}}
Using the asymptotic form of the Bessel function $J_1(kr)$ 
and isolating the outgoing component one easily gets
\eqn\tenthreeone{ |K|^2 = {4\pi \over \omega^3}}
Thus the absorption crosssection is 
\eqn\tenthreetwo{\sigma = |K|^2 |A|^2 = A_H}
where we have used \tentwoseven.  

Thus we get a result similar to that in \unruh , where it was found
that for the 3+1 dimensional Schwarzschild hole the low energy cross
section for scalars is the area of the horizon.  Thus this cross
sections differes by a factor of order unity from the geometrical
cross section, which is the effective cross section expected for
quanta with wavelength much smaller than the horizon size.

The `inner region' that we used above had the form of an infinite
throat with constant diameter, and the solution of the wave equation
in this region was taken as \tentwoone. But this solution assumes that the
length of the throat $L_h$ is much larger than the wavelength
$\omega^{-1}$ of the wave. While this condition is indeed satisfied by
a throat of infinite length, one sees that if the hole is slightly
nonextremal, then the horizon is encountered after a finite length of
the throat, and the calculation must stop at this horizon. In fact in
the calculation of \spenta\ the hole was close to extremality, but the
limit of $\omega\rightarrow 0$ was taken first, so that the length
$L_h$ of the throat became much smaller than the wavelength
$\omega^{-1}$, and the nature of the wave equation and the boundary
condition in the inner region became different from what we had
above. Nevertheless, the cross section computed from this alternative
set of limits agrees with what was computed above (where the limit
of extremality is taken first, and $\omega\rightarrow 0$ is taken
later).

\newsec{Comparison of the D-brane and Classical result}

According to the standard semiclassical derivation of Hawking radiation
\ref\birell{See e.g. N.D. Birrell and P.C.W. Davies, 
{\it Quantum fields in Curved
Space}''(Cambridge University Press, 1982) and references therein}
the total luminosity of particles of a given bosonic species radiated from 
a given black hole in an energy interval $(\omega, \omega + d\omega)$
\eqn\hawone{[{dE (\omega) \over dt}]_{sc} = \sum_{i}{d\omega \over 2\pi}
{\omega\Gamma_{\omega i} \over e^{\beta_H \omega}-1}}
where $i$ collectively denotes the various angular momentum quantum
numbers and $\Gamma_i$ denotes the absorption probability of a wave
with energy $\omega$ and angular quantum numbers $i$.
As mentioned earlier, for scalars the dominant decay at low $\omega$
is into S-waves. In this case $\Gamma_\omega = |A|^2 (\omega)$ computed
in the previous section.

Assuming S-wave dominance and
using \tentwoseven\ in \hawone\ we thus get for our case
\eqn\hawtwo{[{dE (\omega) \over dt}]_{sc} = {A_H \over 8\pi^2}
{\omega^4 d\omega \over e^{\beta_H \omega}-1}}
This is in {\it exact} agreement with the D-brane calculation
\sisix.

\newsec{Discussion}

One interesting feature of the D-brane picture for this five
dimensional black hole is that the low energy excitations on
the brane are scalars or spin-1/2 fermions from the five dimensional
point of view. As a result, in the lowest order process considered
in this paper, photons or gravitons are not produced. Presumably
these may be produced from annihilation of pairs of fermionic open strings
. As mentioned above, in the semiclassical
calculation of Hawking radiation, production of particles with spin one and spin two
are suppressed as well
\foot{We would like to thank J. Maldacena, G. Mandal and L. Susskind for 
discussions on this point}. For four dimensional Schwarzschild holes,
the absorption probability for spin-1 particles behaves as $\sim A_H^2 \omega^4$
 for spin-2 particles it behaves as $\sim A_H^3 \omega^6$, while 
it is   $\sim A_H \omega^2$ for scalars \page . 
It is reasonable to
expect that a similar situation holds for the five dimensional extremal
hole e.g. the absorption probability for spin-1 should be $\sim A_H^2 \omega^6$
and $\sim A_H^3 \omega^9$ for spin-2. These powers of $A_H$ should follow
from the various thermal distribution functions. It is important
to check whether the D-brane results and the semiclassical calculations
for these cases are in agreement as well. We expect to report on
this in the near future.
Another interesting computation is a comparison of the emission of
charged particles (i.e. particles which have a nonzero momentum along
the D-string direction). 

In this paper we have computed the process of emission of scalars from
the collision of open strings on the D-string. It is also possible to
compute the reverse process \dasmathurtwo , where an incoming scalar
interacts with the D-string to create a pair of oppositely moving open
strings on the D-string. This computation leads directly to an
absorption cross section equal to the area of the horizon of the
5-dimensional hole, as expected from the result presented here.

\newsec{Acknowledgements}

We would like to thank A. Dhar, G. Lifschytz, J. Maldacena, G. Mandal
and L. Susskind for discussions. S.R.D. would like to thank the Theory
Group of K.E.K, Theoretical High Energy Physics Group of Brown University,
and Center for Theoretical Physics, M.I.T. for hospitality. 
 S.D.M. is supported in part by 
D.O.E.  cooperative agreement DE-FC02-94ER40818.

\listrefs
\end